\documentclass[twocolumn]{aastex61}

\usepackage{xspace}
\usepackage{upgreek}
\usepackage{color}
\usepackage{natbib}
\usepackage{amsmath}
\usepackage{booktabs}
\usepackage{multirow}
\usepackage{tabularx} 
\usepackage{footnote}
\usepackage{lipsum}
\usepackage[normalem]{ulem}

\shorttitle{}
\shortauthors{Connor}

\begin{document}
\title{Applying deep learning to Fast Radio Burst classification}
\author[0000-0002-7587-6352]{Liam Connor}
\affiliation{ASTRON, Netherlands Institute for Radio Astronomy, 
			Postbus 2, 7990 AA Dwingeloo, The Netherlands}
\affiliation{Anton Pannekoek Institute for Astronomy, 
			University of Amsterdam, Science Park 904, 
			1098 XH Amsterdam, The Netherlands}
\author[0000-0001-8503-6958]{Joeri van Leeuwen}
\affiliation{ASTRON, Netherlands Institute for Radio Astronomy, 
			Postbus 2, 7990 AA Dwingeloo, The Netherlands}
\affiliation{Anton Pannekoek Institute for Astronomy, 
			University of Amsterdam, Science Park 904, 
			1098 XH Amsterdam, The Netherlands}
\shortauthors{Connor \& van Leeuwen}

\email{liam.dean.connor@gmail.com}

\begin{abstract}

Upcoming Fast Radio Burst (FRB) surveys will search $\sim$10\,$^3$ beams on sky with very high duty cycle, generating large numbers of single-pulse candidates. The abundance of false positives presents an intractable problem if candidates are to be inspected by eye, making it a good application for artificial intelligence (AI). We apply deep learning to single pulse classification and develop a hierarchical framework for ranking events by their probability of being true astrophysical transients. We construct a tree-like deep neural network (DNN) that takes multiple or individual data products as input (e.g. dynamic spectra and multi-beam detection information) and trains on them simultaneously. We have built training and test sets using false-positive triggers from real telescopes, along with simulated FRBs, and single pulses from pulsars. Training of the DNN was independently done for two radio telescopes: the CHIME Pathfinder, and Apertif on Westerbork. High accuracy and recall can be achieved with a labelled training set of a few thousand events. Even with high triggering rates, classification can be done very quickly on Graphical Processing Units (GPUs). That speed is essential for selective voltage dumps or issuing real-time VOEvents. Next, we investigate whether dedispersion back-ends could be completely replaced by a real-time DNN classifier. It is shown that a single forward propagation through a moderate convolutional network could be faster than brute-force dedispersion; but the low signal-to-noise per pixel makes such a classifier sub-optimal for this problem. Real-time automated classification may prove useful for bright, unexpected signals, both now and in the era of radio astronomy when data volumes and the searchable parameter spaces further outgrow our ability to manually inspect the data, such as for SKA and ngVLA.

\end{abstract}

\section{Introduction}

Fast radio bursts (FRBs) are bright, millisecond-duration, extragalactic 
radio transients, characterized by dispersion measures (DMs) that are significantly 
larger than the expected Milky Way contribution. They have been 
detected at flux densities between tens of micro Janksys to 
tens of Janskys \citep{lorimer07, thornton-2013, petroff2015, ravi2016}. 
The majority of early detections 
were made with the Parkes telescope 
multi-beam receiver,
but in recent years detections have been made at Arecibo \citep{spitler2014},
Green Bank Telescope (GBT) \citep{masui-2015b}, 
the Upgraded Molonglo Synthesis Telescope (UTMOST)
\citep{caleb2017}, and 
the Australian Square Kilometre Array Pathfinder (ASKAP) \citep{bannister2017}.  
FRB 121102 is the only source known to repeat 
\citep{2016Natur.531..202S, 2016ApJ...833..177S}, 
allowing for the first host galaxy 
localization using very long baseline interferometry 
(VLBI)
\citep{2017ApJ...834L...8M, 2017ApJ...834L...7T}. 
Recently, the repeating bursts from this source were
found to be almost 100$\%$ linearly 
polarized with a Faraday rotation measure (RM) of 
$10^5$\,rad\,m$^{-2}$ \citep{michilli2018}. 

There are likely thousands of detectable events each day 
across the full sky, but only $\sim$\,50 have been observed to-date. 
This is due to the moderate field of view (FoV) and relatively 
low duty cycle of current FRB surveys. Still, such surveys have 
produced thousands of false-positive triggers for each true FRB, 
the diagnostic plots of which have traditionally been inspected by eye 
\citep{masui-2015b, connor2017, caleb2017, foster2018}. 
For upcoming fast 
transient surveys, the false-positive problem will be intractable if 
single-pulse candidates are to be human inspected, even 
with rigorous removal of radio frequency interference (RFI). 
The Canadian Hydrogen Intensity Mapping 
Experiment (CHIME) will search 1024 beams 
at all times, between 400--800\,MHz and up to 
very high DMs \citep{ng2017}. 
The Aperture Tile in Focus (Apertif) experiment 
on the Westerbork telescope
will continuously search thousands of synthesized beams 
at 1.4\,GHz \citep{leeu14}. 
ASKAP
\citep{bannister2017} and UTMOST \citep{caleb2017} are also 
expected to have high detection rates, searching many beams 
with high duty-cycle. As a result, we will go from
roughly five new FRB detections per year (2012--2017) to, potentially,
thousands ($>$\,2019). This will also correspond with an orders-of-magnitude increase 
in the number of false positive candidates, meaning the generation 
of such events must be mitigated, and the process of sifting through 
them must be automated.

\begin{figure*}
	\centering
	\includegraphics[width=\textwidth]{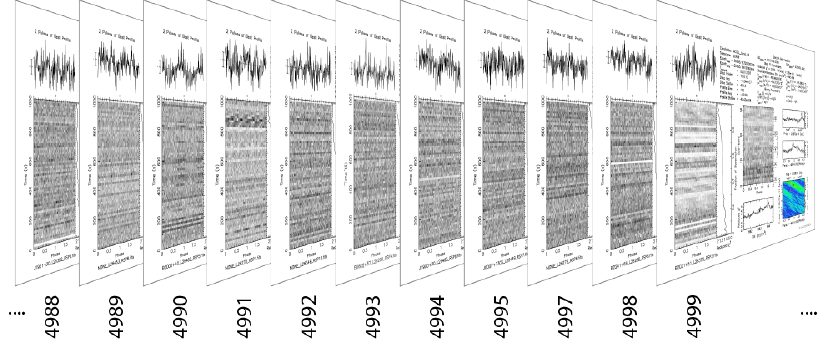}
	\caption{A small subset of the real-life diagnostic data used in the LOTAAS survey with LOFAR
          \citep{2014A&A...570A..60C}. Out of $\sim$20,000 candidates, pulsar J0613+3731 was found by eye in plot
          \#4993.}
\label{fig-example}
\end{figure*}

In pulsar searching, the problem is arguably worse
due to the larger number of parameters involved, 
like period and its derivatives.
Over the last decades, the ranking of pulsar 
candidates has involved an initial step of selection through simple heuristics,
the main one being the peak signal to noise ratio of the profile over the noise. 
Thereafter, the astronomers go through the ordered list of candidate plots, looking for further pulsar signs such as
broad-band, properly dispersed signal; a sharply peaked (not sinusoidal) folded profile; and steady emission throughout the
observation. An experienced pulsar astronomer can average 1--2 plots per second, and human brains are very capable of
singling out the most promising candidates.
But modern multi-beam pulsar surveys, and the increasing bandwidths and new frequencies outside of radio-quiet protected
spectrum are making this approach unfeasible. 
A telescope like LOFAR employs many hundreds of beams \citep{ls10}, and produces vast numbers of candidates.
The LOFAR pilot surveys LPPS and LOTAAS \citep{2014A&A...570A..60C} produced $\sim$20,000 candidates, that were ranked
and perused by humans. This took about four person-days. It found the first two pulsars with LOFAR.
Shown in Fig.~\ref{fig-example} is a subsection of the ranked list that included pulsar J0613+3731.

This approach is, however, reaching the limits of what is efficient. For a long-integration, multi-beam LOFAR search for
young pulsars in supernova remnants, the lead author of \citet{sl18} checked, by eye, the staggering number of
140,000 periodic candidates 
plus about 15,000 single-pulse candidates.
This amounted to three full-time person weeks of time.

While efforts like the Pulsar Search Collaboratory \citep{2013ApJ...768...85R} have been successful in 
engaging hundreds of citizen scientists in ranking and analyzing candidates from GBT pulsar survey data, the overall
person-power requirement remains unchanged and daunting.

The necessity 
of replacing manual inspection
has led to a variety of approaches. 
\citet{zhu2014} developed a sophisticated 
framework for pulsar candidate ranking, 
using multiple machine learning techniques to emulate 
a human expert inspecting diagnostic plots for tens of 
thousands of pulsar candidates from PALFA. 
They used convolutional neural networks (CNN) in tandem with 
support vector machines (SVM)
on the pulsar candidates' two-dimensional arrays, and ANNs 
with SVMs on one-dimensional data products, like pulse profile. 
Next, \citet{guo2017} utilized a 
convolution generative adversarial network (DCGAN) to improve 
the ability of deep CNN classifiers. 

The LOFAR Tied-Array All-Sky Survey (LOTAAS) uses 222 digitally formed tied-array beams per
pointing, and its search pipeline reports the $\sim$100 best periodic candidates per beam.
Currently, in early 2018, 1500 pointings have been observed, and over 30 million periodic candidate signals were
found. These can clearly no longer be inspected by eye. 
Thus \citet{2016MNRAS.459.1104L} built a tree-based machine learning classifier, 
using a set of features from these periodic candidates. Using the first LOTAAS data, \citet{2018MNRAS.474.4571T} next
improved the feature selection, increased the training set size, and combined 5 decision trees into an ensemble
classifier to further enhance the algorithm recall.
Overall, these periodicity classifiers have helped discover many tens of new LOFAR pulsars.  

The application of artificial intelligence (AI) 
to \emph{single pulse} classification
is less well developed, in part due to the nascency of FRB and rotating radio transient (RRAT) 
science. Though there is significant overlap with candidate ranking 
in pulsar periodicity searches, the problem of single-pulse 
classification has several distinctions, particularly for upcoming 
multi-beam real-time FRB surveys. These include the need for real-time 
classification for VOEvents and voltage dumps, as 
well as the usefulness of multi-beam information. 
\citet{devine2016} developed a method for identifying 
clustered groups of dispersed pulses, primarily in order to discover 
pulsars that might be missed by periodicity searches. There, 16 group features (e.g., start-end DM, maximum signal-to-noise ratio S/N).
are used in six traditional machine learning algorithms 
to find the best combination of hyperparameters and classifier. 
In Arecibo's commensal FRB search, ALFABURST, \citet{foster2018} 
built a training set on 15,000 events and extracted 409 features 
from each. A random forest was then applied to group each trigger into 
one of nine classes. 
For the LOTAAS survey on LOFAR, \citet{mhb+18}
adapted the Gaussian-Hellinger Very Fast Decision Tree used for periodicity classification \citep{2016MNRAS.459.1104L},
and implemented a single-pulse search pipeline.
It was trained on $3.5\times10^4$ labelled RFI instances and 
$1.8\times10^4$ thousand single pulses from 
47 known pulsars as recorded in the LOTAAS data, 
and has discovered 7 pulsars based on features like 
pulse width, DM, and S/N vs. DM \citep{mhb+18}.

A next step, and challenge, in wide-field FRB searching will be the ALERT\footnote{\url{www.alert.eu}} survey on Apertif.
An hierarchical series of beamforming starts with 39 compound beams (cf.~Fig.~\ref{fig-beams}) formed on each of the
phased array feeds in the 12 dishes equipped with these. Every compound beam is next coherently beam-formed in 12 offset grating response beams; a refinement step of on-the-fly beam-forming, for removing chromatic sidelobe effects within this wide-bandwidth system, finally increase the beam count by a factor 6 for a total of $\sim$2800 synthesised beams \citep{2017arXiv170906104M}. These are searched in a real-time 
single-pulse pipeline
powered by a large Graphical Processing Unit (GPU) cluster \citep[ARTS;][]{artsso18}.
The ALERT survey will run 24/7
for approximately 3 calendar years. At 2$\times$ the number of
beams, 5$\times$ the bandwidth,  and more than 10$\times$ the on-sky time of LOTAAS, the number of single-pulse
candidates produced in ALERT is expected to not be humanly manageable.

In this paper we apply deep learning to the problem of 
single-pulse classification, for the first time. 
We develop a flexible toolkit 
that allows for the construction of 
hierarchical deep neural networks with multiple data products  
as inputs. Our approach will be useful both for multi-beam 
surveys as well as single-pixel telescopes. Classification 
can be done very quickly on GPUs by using the highly optimized 
software library, {\tt TensorFlow}, 
which will be necessary if post-dedispersion 
real-time decisions are to be made. 
The paper is organized as follows: In Sec.~\ref{sect-dl}
we introduce the key concepts of deep learning, and discuss its 
advantages of more traditional machine learning algorithms. 
In Sec.~\ref{sect-arch} we describe our model's tree-like architecture, 
and show how arbitrary data inputs and feature extraction branches 
can be added to the network. We also offer tools to remedy the black
box problem. 
Sec.~\ref{sect-data} discusses how we assemble a 
labelled training set, despite there being only two dozen FRBs
to-date. We then present the classifier's results in Sec.~\ref{sect-results}, 
showing that very high recall and accuracy can be achieved 
with a sufficiently comprehensive training set. 
Sec.~\ref{sect-realtime} asks if it would be possible 
to replace real-time dedispersion backends with a neural network 
classifier. We show that, somewhat surprisingly, forward-propagation through 
a simple convolutional neural network could be faster than 
brute-force dedispersion. However, simpler statistical approaches, 
like current dedispersion algorithms will always be more optimal, 
so such AI-based real-time classification may only be useful 
once unsupervised deep learning is more developed as a field.

\section{Deep learning}
\label{sect-dl}

Within the concentric circles of artificial intelligence (AI),
machine learning has made the most progress in recent decades.
Machine learning refers to a class of tools that aims to
let computers learn without being explicitly programmed.
Representation learning is a further subset of machine learning whose goal 
is not only to model the mapping from input features to output, but 
to actually discover the feature (or representation) itself 
\citep{Goodfellow-et-al-2016}.
Representation learning circumvents the limitations of 
``feature engineering'', in which data-specific features must 
be chosen by hand. This often requires domain expertise and 
can be time consuming. With real world data, extracting the salient 
features from input data tends to be difficult. 
The last subset in these concentric circles, deep learning, 
helps with the representation problem by building complexity out 
of multiple, smaller, representations
\citep{lecun2015, Goodfellow-et-al-2016}.

\begin{figure}
	\centering
          \includegraphics[clip,trim=3in 1in 3in 1in,width=\columnwidth]{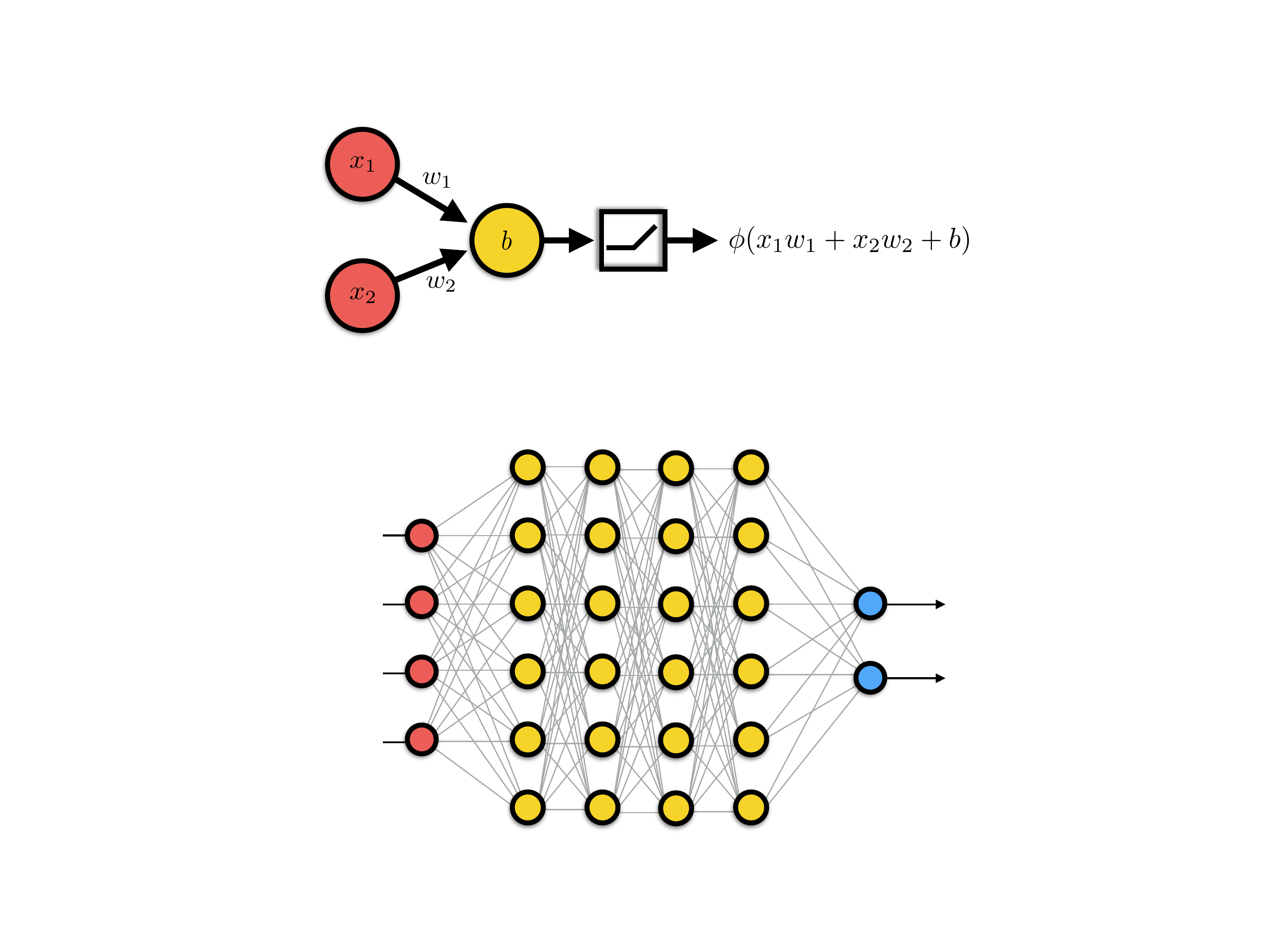}
	\caption{A schematic diagram of a perceptron (top), and a collection 
			 of perceptrons combined to form a neural network (bottom). 
			 The yellow nodes are artificial neurons, the red are input data, and the 
			 blue are output classes. The perceptron's output is a non-linear function 
			 of the input vector, $\mathbf{x}$, projected onto the 
			 weight vector, $\mathbf{w}$, with some offset $b$. The network 
			 shown is a ``deep neural network" because it has multiple hidden layers.}
\label{fig-perceptron}
\end{figure}

A deep neural network (DNN) is typically just a neural network 
that has multiple hidden layers.
DNNs make use of the ``multilayer perceptron'', 
a combination of artificial neurons and connections between them. 
Each subsequent hidden layers represents higher levels 
of abstraction of the input. A cartoon example of such a 
network is shown in Fig.~\ref{fig-perceptron}. 
The perceptron has weights 
$\mathbf{w} = \left ( w_1, .., w_n \right )$
corresponding to each of the $n$ connections with 
between the input data, $\mathbf{x}$, and a neuron, as well as a single 
offset value, $b$. 

The perceptron computes 
a linear combination of the input with the weights, such that 

\begin{equation}
z = \mathbf{x}\cdot\mathbf{w} + b.
\end{equation}

\noindent A non-linear activation function 
is then applied to $z$, such that the output of 
the perceptron is some function $\phi(z)$. 
Such activators must be non-linear; otherwise, no 
matter how many hidden layers are in a network, the output 
would simply be a linear function of the input. It is also a problem 
that a linear activator's gradient is independent of the input, making 
training via gradient descent impossible.
Common examples include the logistic function, 

\begin{equation}
\phi(z) = \frac{1}{1 + e^{-z}},
\end{equation}

\noindent a hyperbolic tangent, 

\begin{equation}
\phi(z) = \mathrm{tanh}(z),
\end{equation}

\noindent or a rectified linear unit (ReLu),

\begin{equation}
\phi(z) = \left\{\begin{matrix}
z \,\,\,\mathrm{for}\,\,z \geq 0 \\
\,\, 0 \,\,\,\mathrm{for}\,\,z \leq 0 \,\,.
\end{matrix}\right. 
\end{equation}

\noindent In this work we mostly use ReLu functions, 
which have been empirically found to be highly effective 
\citep{maas2013rectifier}.

It is easy to see 
how complex functions could, in principle,
be modelled by finding the right weight and offset 
parameters for each connection and neuron, 
given enough labelled input data. What is less obvious is why 
this can be done with relatively few parameters. The size of 
the input space of a 100$\times$100 greyscale image is 
$256^{10000}$, yet in many cases the mapping from 
input image to output class can be well approximated 
by $\sim$\,millions of parameters \citep{lin2017}.

Irrespective of \textit{why} deep learning has 
been so successful, its advancement of AI in recent years 
is undeniable. Deep learning has 
led to quantum leaps in self-driving car AI, 
super-human image recognition, 
natural language processing, and machine translation 
\citep{lecun2015, goldberg2015, Goodfellow-et-al-2016}. 

One type of network, the convolutional neural network (CNN), 
has proven particularly powerful. 
Such architectures use convolution along with pooling 
to extract high-level features from input data. 
In the case of image recognition, an input image 
is typically convolved with multiple different kernels, 
which are meant to find structure in the data 
and identify relevant attributes. 
A non-linear activation function 
is then applied to the multiple convolved images, or ``feature maps''. 
Each convolution step is followed by a ``pooling'' layer. 
The pooling can be as simple as taking the maximum pixel value 
in a small region (max pooling), 
and is meant to act as a summary statistic, allowing 
for some translation invariance and robustness against noise 
\citep{Goodfellow-et-al-2016}.
An example of our CNN is shown in Fig~\ref{fig-CNN2d} with 
the real activations of an input dynamic spectrum generated 
by a trained model.

\section{Multi-input CNN}
\label{sect-arch}

The classification of dedispersed single-pulse 
candidates is slightly different from other problems to 
which CNNs have been applied. For example,  
training a model for image recognition of, say, 
different breeds of dogs, requires 
building a network that can learn a very large and 
complex image space 
based on photos with effectively infinite S/N per pixel.
FRBs occupy a much smaller 
volume of image space, but S/N per pixel is $\sim$\,1. 
This turns out to be considerably less difficult 
than some other applications.
Therefore we can achieve high recall and precision 
with modest-sized training sets (tens of thousands of triggers)
and relatively few layers. 

\subsection{Frequency-time data}
The most informative input data array is the frequency-time 
intensity data, or dynamic spectrum. This input lends itself well to 
a 2D CNN, where image topology is preserved.
In other machine learning algorithms such as support 
vector machines, 2D input data are flattened 
into a 1D vector.  
Dedispersion algorithms search a frequency-collapsed time series, 
triggering on outliers in one dimension. Therefore, at
a single DM, valuable spectral information is thrown out where most 
false positives from thermal noise will not look like a 
broad-band pulse. By applying a deep CNN to the dynamic spectrum 
image the model can discriminate based on frequency structure. 

Our dynamic spectrum model is a CNN with two convolutional layers, two 
max-pooling layers, and two fully connected layers. 
We pre-process input data by demanding that each trigger 
have unit variance and zero median. We find that 
input frequency-time arrays of shape $32\times64$ allow 
for sufficient signal per pixel, but there is flexibility in 
the resolution of the input image.

\begin{figure*}
\includegraphics[width=\textwidth]{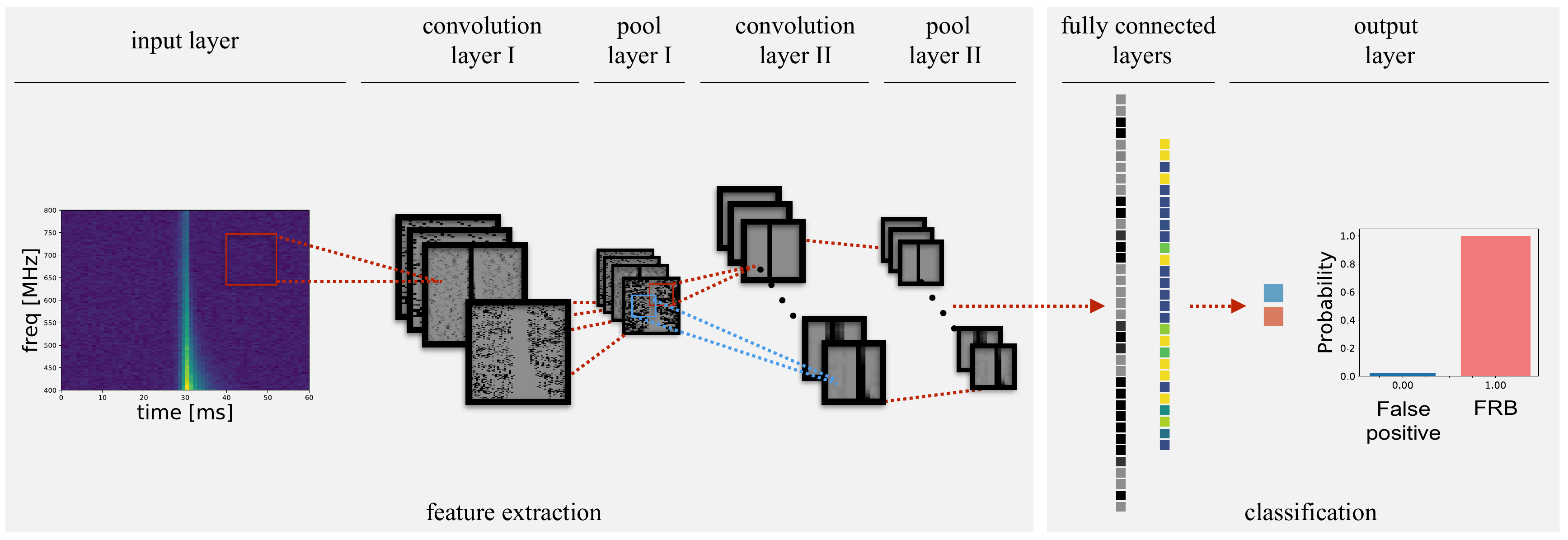}
	\caption{An example of our CNN architecture for the frequency-time array 
		     of a dedispersed FRB. The input image is a simulated strong, 
		     scattered burst, artificially bright for the sake of this example.
		     The input data are convolved with 32 different convolutional kernels, the results 
		     of which are fed through a non-linear ReLu function, 
             then reduced in a max pooling layer. 64 more kernels 
		     are then applied to the binned arrays, and pooling is done again by 
		     taking the maximum value in each 2$\times$2 bin. Each 
		     successive layer in the network's feature extraction 
		     component is meant to discover features  
		     at higher levels of abstraction. We have included in this figure 
		     the real actualizations of a trained model for this input image. 
		     These give an idea of what the neural network's classification is based on, 
		     and can help with the `black-box' problem. 
		     After the fully 
		     connected layers, a probability is assigned to the trigger's likelihood of 
		     being an FRB.}
\vspace{0.25 in}
\label{fig-CNN2d}
\end{figure*}

A scaled-down 
example of this architecture is shown in Fig~\ref{fig-CNN2d}.
The figure allows not only 
for visualization of the network's 
architecture, but also a way of peeking inside the model and 
looking at each hidden layer's activations. By saving the trained 
network's weights and convolutional kernels, a given input array can be 
forward-propagated through the model to produce activations 
at each layer. The activations give one an idea of what the neural 
network ``sees'' in a given hidden layer, which alleviates the black box 
problem of DNNs, and is also helpful as a debugging tool. 
The CNN clearly tries to separate 
the input data's background noise from the
features intrinsic to the FRB pulse, such as a scattering tail.

\subsection{DM-time data}

The DM-transformed data is a DM-time array whose 
rows are the frequency-collapsed time stream 
at a given DM. Broad-band, dispersed pulses will 
show up as a small island of preferred DM-time pairs,
exhibiting a bow-tie pattern due to a degeneracy between 
optimal dispersion measure and pulse arrival time. 
For this input we also use a simple 2D CNN, 
with DM-time arrays of dimension $100\times64$. An example 
is shown in the third row from the top of Fig.~\ref{fig-tree}

\subsection{Pulse profile}

We apply a one-dimensional CNN to the pulse 
profile dedispersed to the DM that maximizes S/N. 
The DNN's first convolutional layer applies 32 
length-5 kernels with strides of 2. After a 
1d-max-pooling, another convolutional layer is applied 
with 64 length-2 kernels. The output is flattened 
and applied to a fully-connected layer with 
1024 neurons. 

\subsection{Multi-beam detections}
Most upcoming competitive FRB surveys will search multiple beams 
simultaneously. Objects beyond an antenna's far-field limit 
are not expected to be seen in more than a couple of adjacent beams, 
whereas terrestrial RFI can be detected in many non-neighbouring beams. 
Other groups have taken this into account by rejecting triggers 
that showed up in unexpected beam permutations.

We allow our model to learn such permutations without explicitly 
telling it a given telescope's on-sky beam configuration. 
This was done using a simple feed-forward neural network 
whose input data is a 1D length-$N_{\rm beam}$ vector of detected S/N
per beam. If no event was found above the cutoff significance, 
a S/N of zero is assigned. After training, the model learns 
which beams ought not to trigger simultaneously, and which 
combinations are acceptable for a real astronomical detection.

\subsection{DNN tree}

\begin{figure*}
\includegraphics[width=\textwidth]{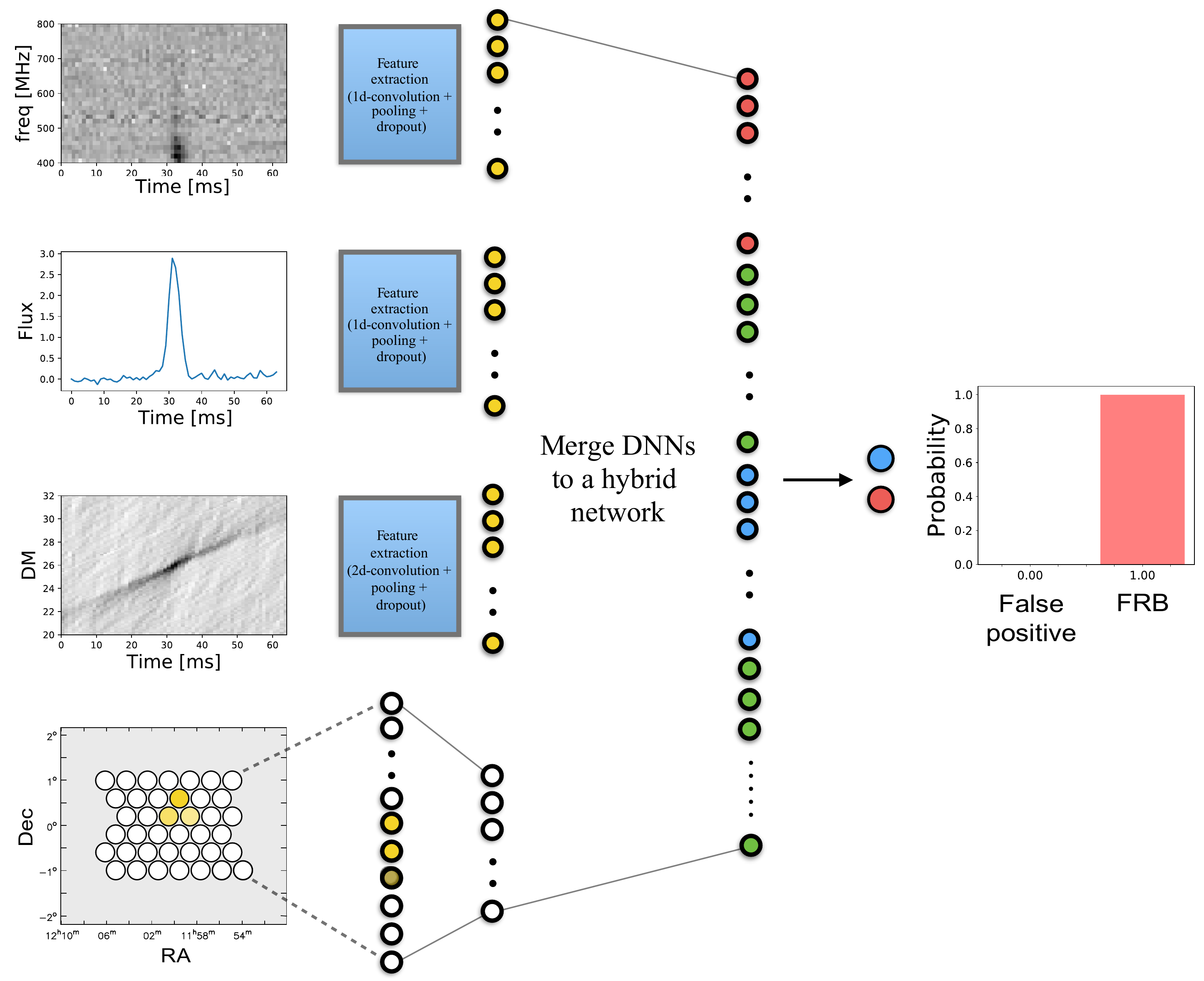}
	\vspace{0.5 cm}
	\caption{A hierarchical hybrid neural network built from 
			 concatenating multiple nets after their feature 
			 extraction layers, creating a large fully-connected layer
			 resulting in a single binary classification for all inputs. 
			 Here we use as inputs the dedispersed 
			 frequency-time intensity array, the frequency-collapsed 
			 pulse profile, DM-time array, and multi-beam detection S/N,
			 but our framework allows for any combination of these 
			 as well as additional input data products and networks. Since 
			 the full, merged DNN is trained together, the model 
			 will learn the relative importance of each input---for example,
			 the dedispersed intensity array will tend to have more 
			 predictive power than the multi-beam statistics, thus the former
			 will to have a greater influence on the output probability.}
\label{fig-tree}
\end{figure*}

We developed a multi-input neural network, to which arbitrary additional nets
can be appended. The idea is to extract features from each input data product 
independently, since a given burst's salient characteristics will depend 
on the space in which it is being viewed. The multiple networks can then be concatenated 
at the classification layers (in our case fully connected layers after convolution), creating a 
hierarchical tree-like neural network, shown in Fig.~\ref{fig-tree}.

The first three data products we use in Fig.~\ref{fig-tree} are not independent. 
Indeed, their information content is highly redundant: 
The 1D pulse-profile is simply the dedispersed frequency-time 
array collapsed along the frequency axis; and the DM-time array is the frequency-time  
data after the DM transform.
And yet empirically, better results are achieved by including combinations of 
the three than any individual one. This is because the feature extraction step 
is imperfect, so projecting the data in different ways allows the networks 
to detect different modes. The same is true for human classifiers. When sifting 
through pulse candidates one often looks at multiple statistics and figures 
with overlapping information. 

\section{Training data}
\label{sect-data}

The two conditions that have allowed deep learning 
to thrive in this decade have been the availability of 
large, labelled data sets, and computers that can 
train multilayer models in a reasonable amount of time. 
Despite the high all-sky event rate of FRBs, only a couple 
of dozen events have been discovered to-date 
(\citealt{petrofffrbcat}\,\footnote{\url{http://www.frbcat.org}}).
This presents a problem that does not exist for 
pulsar candidate classification. The small catalogue of real events 
is probably not yet a representative sample of the underlying burst 
population, nor is it big enough to build a meaningful training 
set for machine learning, deep or otherwise. 

This means bursts can either 
be simulated, or single pulses from Galactic pulsars 
could be used as an approximation, or 
a combination of both. In this work we choose to simulate most of
our ``true-positives'' and use false-positives that have been generated 
in real surveys and labelled by eye. 
We do not use single pulses from Galactic pulsars as a primary training 
set for the following reasons: Even though a large number of pulses can 
be collected from individual pulsars,  the variation 
within FRBs \citep[cf.~the ASKAP set;][]{bannister2018} appears to be larger than the pulse-to-pulse variation from 
a single pulsar, in terms of pulse characteristics 
(width, scattering, frequency structure, etc.). 
The differences between FRBs could also be larger than the
variation between Galactic pulsars, given 
the extreme conditions they appear to live in \citep{masui-2015b, michilli2018}.
FRB 121102 shows frequency structure 
on at least two different scales, and has bursts 
ranging in duration from 30\,$\mu s$
to several ms \citep{michilli2018}.
Therefore, while a considerable training set could be built up 
from Galactic sources, the resulting model might be over-fit to the properties 
of the pulsars whose single pulses are bright enough to detect. Finally, 
while de-dispersed FRBs are qualitatively similar to single pulses 
from nearby pulsars ($\sim$\,millisecond-duration, broad-band, etc.), 
there may be systematic differences that are not obvious or visible, but that would bias the learner. In a simulated set there is more control and insight into the parameters producing the set that we train against.

While we choose to simulate our true-positives, 
false-positive triggers should not be simulated.  
Events generated by RFI, thermal noise, 
or dropped packets, occupy a much large volume of image space 
than single-pulses from FRBs, RRATs, or pulsars. 
Simulating RFI triggers would be difficult since there is 
no good model that describes such events. On top of that, each 
instrument will produce a different set of false-positives 
due to their disparate RFI environments and signal-processing back-ends. 
Conversely, single-pulses from FRBs can be modelled 
with far fewer parameters. Though they can suffer to varying degrees 
from temporal scattering, 
frequency scintillation, and DM-smearing, these effects can, in principle, 
be accounted for. By including a large collection of events 
that plausibly samples the full phase space of fast radio bursts
in one's training set, 
a sufficiently sized neural network
can learn to identify a wide range of pulses. Casting such a 
a wide net should catch true single-pulses. 

We have built a training set from the 1268 hours of data in the 
CHIME Pathfinder incoherent-beam FRB search,
plus simulated events injected into those data \citep{connor2017}. 
We use 4650 events that triggered the dedispersion pipeline with a 
S/N above 10, but were found to be false-positives after 
inspection by eye. We then inject 
an equal number of simulated FRBs
drawn from the distributions described in Sec.~\ref{sect-sims}.
Single pulses from known Galactic pulsars 
also triggered the search pipeline, including Crab giant pulses 
and individual pulses from PSR~B0329+54. These astronomical true-positives 
were separated and used later in the verification of our model.
For our Apertif model, the training set consists of 21246 
candidates, half of which are known false positives. Of 
the remaining triggers, roughly 9800 are simulated FRBs 
added to real data, along with a couple of 
hundred single pulses from Galactic pulsars.

\begin{figure}
\center
   \includegraphics[clip, trim={2.1in, 0.1in, 2in, 0.5in}, width=\columnwidth]{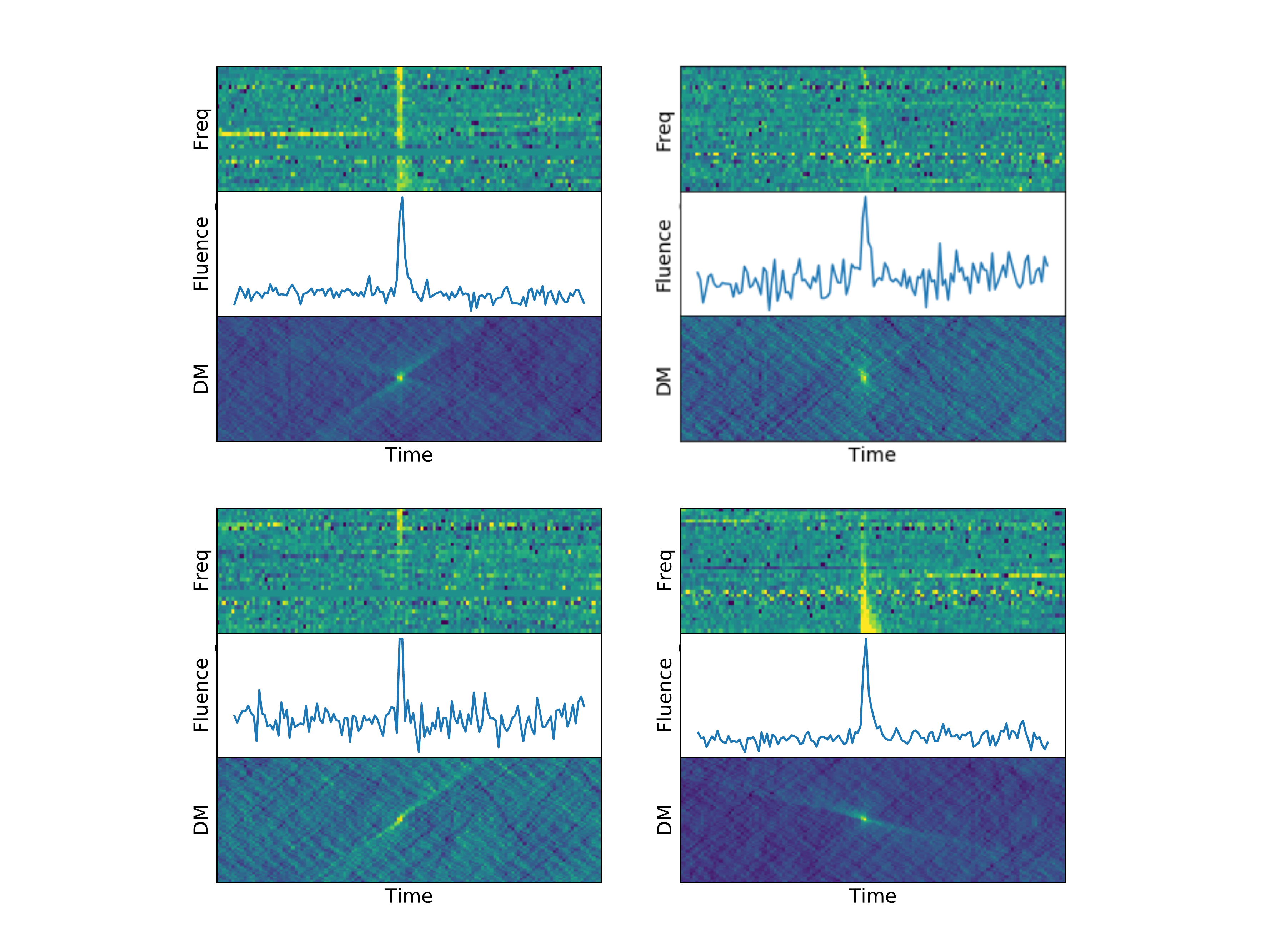}
	\caption{Four examples of simulated FRBs injected into real data.
			 By combining thousands of these true-positives with known 
			 false-positives, we have built a large labelled training set.
			 The three panels in each sub-figure are the frequency-time 
			 intensity array of the dedispersed pulse, the frequency-averaged 
			 pulse profile, and the DM-time intensity array (top to bottom). Combinations 
			 of these data products can be used as inputs to the multi-input 
			 neural net described in Sec.~\ref{sect-arch}.}
\label{fig-triggers}
\end{figure}

\subsection{Simulation}
\label{sect-sims}

We simulate FRBs in one of three ways. 
The preferred approach is 
to randomly inject events in the data using the 
real-time tree dedispersion pipeline 
{\tt burst\_search}\footnote{\url{https://github.com/kiyo-masui/burst\_search}}. 
Another way is to add simulated FRBs to real background data
that has already been dedispersed to a random DM. Finally, 
we can add pre-dedispersed FRBs to gaussian noise.  

We calculate the pulse profile at each frequency 
by convolving a gaussian with a 
scattering profile, 

\begin{equation}
s(t) = \frac{1}{\tau_\nu} e^{-t / \tau_\nu}
\end{equation}

\noindent where $\tau_\nu$ is the 
scattering timescale at a frequency $\nu$ and is given by,

\begin{equation}
\tau_\nu = \tau_0 \left( \frac{\nu}{\nu_{\rm ref}}\right)^{-4},
\end{equation}

\noindent for a reference frequency $\nu_\mathrm{ref}$.
The gaussian is taken to have width $t_I$, 

\begin{equation}
t^2_I = t^2_i + t^2_{\textrm{samp}} + t^2_{\textrm{DM}}
\label{eqn:smearing}
\end{equation}

\noindent where $t_i$ is the intrinsic pulse width, 
$t_{\textrm{samp}}$ is the sampling time, and 
$t_{\textrm{DM}}$ is the DM-smearing timescale. 

Burst fluence is drawn from a Euclidean distribution, 
resulting in a S/N distribution of pulses that is also approximately
Euclidean. Pulse widths are assumed to follow a log-normal 
distribution with mean 1.6\,ms, resulting in 
widths between 0.1--50\,ms. Scattering measure is log-uniform, 
in a way that roughly one in five bursts is 
noticeably temporarily scattered. Frequency scintillation is included via 
intensity modulation across the band using the positive 
half of a sinusoid with random phase
and random decorrelation bandwidth. The scintillation bandwidth distribution 
is such that only about one third of simulated bursts show discernible 
frequency variation, consistent with the current population 
of detected FRBs. We take a uniform distribution of spectral index, $\gamma$, 
between -4 and +4, where $F_{\nu} \propto \nu^\gamma$.

After signals are injected, events are kept if their 
S/N falls between 8--80. Ultra-bright events are 
discarded because they do not add much predictive 
power to the trained model; if a 500\,$\sigma$ event 
is found in the data, a model that has learned 80\,$\sigma$
events will still find it.
All data are preprocessed to have unit variance and zero median.
Uniformity in the treatment of both the simulated FRBs and 
the detected false-positives is important, because otherwise 
the binary classifier will learn based on trivial differences 
like noise RMS or power offset. In Fig.~\ref{fig-triggers} 
we show examples of FRBs generated.
The exact parameters of the distributions chosen are, of course, 
tunable, and will be subject to change depending on the 
survey for which the model is being built. The simulation
tools are availabel on github\footnote{\url{https://github.com/liamconnor/single\_pulse\_ml}}.

\subsection{RFI excision vs. classification}

RFI will be an appreciable problem for all 
upcoming FRB surveys.
Though we may want to mitigate RFI as much as 
possible, there exists a trade off between 
pre-dedispersion RFI cleaning and 
false positive rejection post-triggering. If 
one wants to minimize the number of RFI events triggered 
by a dedispersion algorithm, then data preprocessing needs to 
be done thoroughly. However, this runs the risk of over-cleaning 
the data and removing astronomical events. For example, if 
one of the steps in the RFI excision pipeline is a standard sigma-cut 
in which samples above, say, $3$\,$\sigma$ in their local neighborhood are removed, then 
events like the Lorimer burst \citep{lorimer07} or FRB 150807 \citep{ravi2016}
could be missed, especially if the events fluctuate in frequency and time. 
Another approach would be to preserve as many triggers as 
possible by doing modest or no RFI cleaning, allow large numbers 
of false positives to trigger the dedispersion pipeline, and to only 
make a final decision after the triggers have been classified by 
the machine learning algorithm. This would make sense if 
one had high confidence in one's classifier, otherwise real 
signals might drown in the flood of false positives. 
The solution is probably somewhere in between the two extremes:
Data ought to be cleaned enough that the RFI within a time-frequency 
block containing an FRB does not decrease the event's S/N. And a balance 
must be struck between the number of triggers generated and the risk 
of missed events, i.e. false negatives, which will require experimentation
and will be survey-dependent. 

\section{Results}
\label{sect-results}

In order to assess our model's performance, 
we use standard metrics based on the confusion 
matrix.
``accuracy'' corresponds to the fraction of 
classified events that were labelled correctly,
``precision'' is the ratio 
of true positives to the number of events 
classified as positives, and ``recall'' is the fraction 
of true events that were labelled as such. Using 
TP, TN, FP, and FN as the number of true positives,
true negatives, false positives, and false negatives 
respectively, the metrics are given by,

\begin{equation}
\rm acc = \frac{TP + TN}{TP + TN + FP + FN}
\end{equation}

\begin{equation}
\rm precision = \frac{TP}{TP + FP}
\end{equation}

\begin{equation}
\rm recall = \frac{TP}{TP + FN}.
\end{equation}

\noindent We care about the recall rate, because 
this determines the probability of missing an FRB. 
However, the precision is also important in case of real-time 
triggering. If voltage data are to be written after 
the dedispersion pipeline is triggered, as with ASKAP 
or UTMOST, then one must be sure that most of those 
events really are FRBs. The same is true for email notifications, 
VOEvents  \citep{petroff2017VO}, such as when Apertif will trigger 
LOFAR's transient buffer boards for low-frequency localization.

\begin{figure}
\center
\includegraphics[clip,trim={0.2in, 0.2in, 0.2in, 0.28in}, width=\columnwidth]{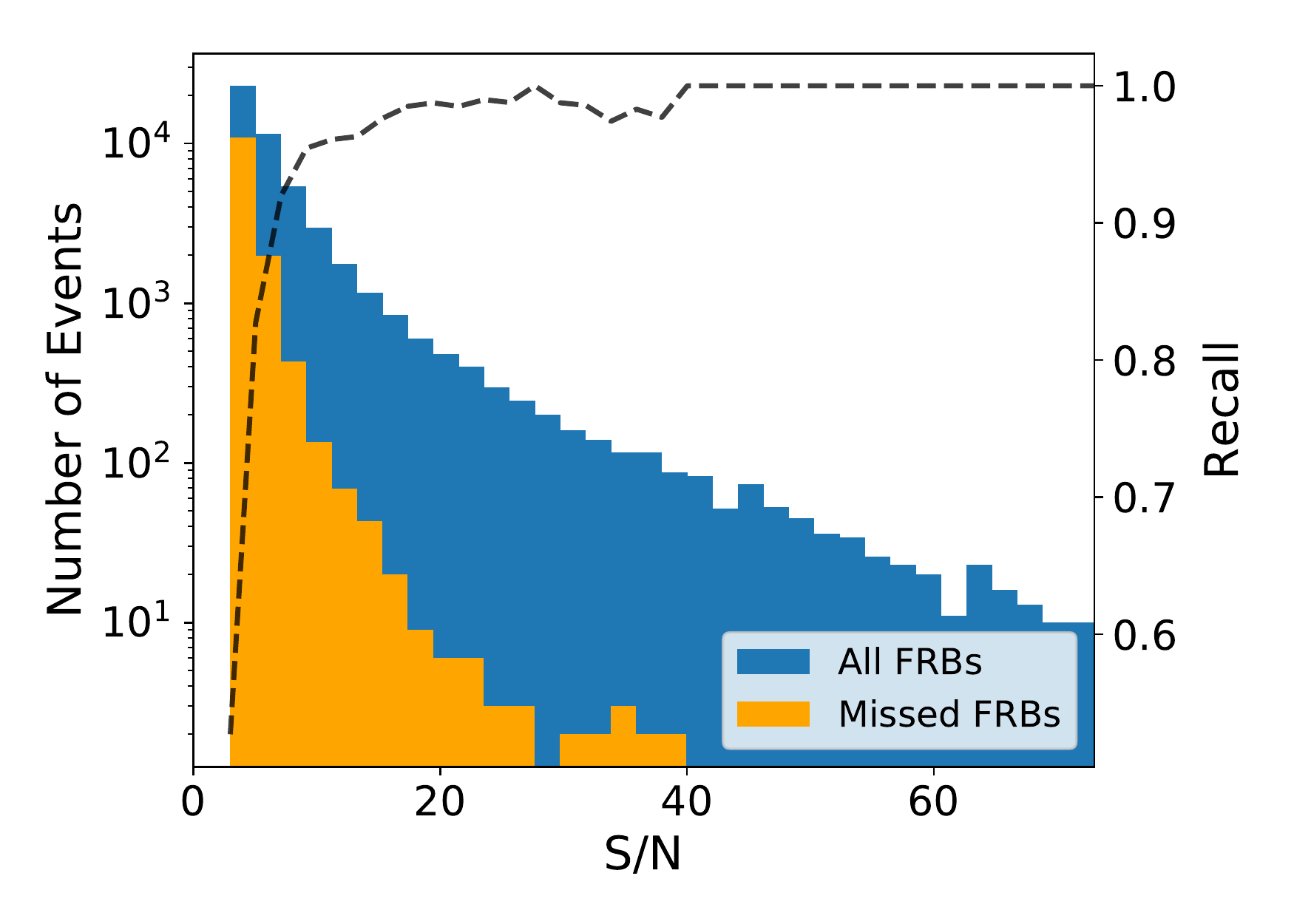}
	\caption{Statistics of missed FRBs as a function of signal to noise. 
			 The histogram shows the distribution of 50,000 simulated FRBs 
			 in the test set (blue) as well as the events from that test 
			 set that were mislabelled as RFI by our frequency-time 2D CNN (orange). The
			 false-negative rate goes to 0.5 for low S/N, as expected, 
			 since a binary classifier with no predictive power will classify 
			 correctly half of the time. 
			 The fraction of recovered 
			 events, or recall, gets close to 1 for high S/N.  
	}
\label{fig-signoise}
\end{figure}

In Fig.~\ref{fig-signoise} we plot recall as a function 
of S/N. As expected, at very low S/N the model loses 
its predictive power: Recall drops to 50$\%$ because 
the algorithm is making a random guess at binary 
classification. However, above $\sim$\,8\,$\sigma$ the 
fraction of missed FRBs is quite flat, and also low, 
with recall and accuracy above 99$\%$.

\subsection{Cross-validation with pulsars}

\begin{figure}
\includegraphics[width=\columnwidth]{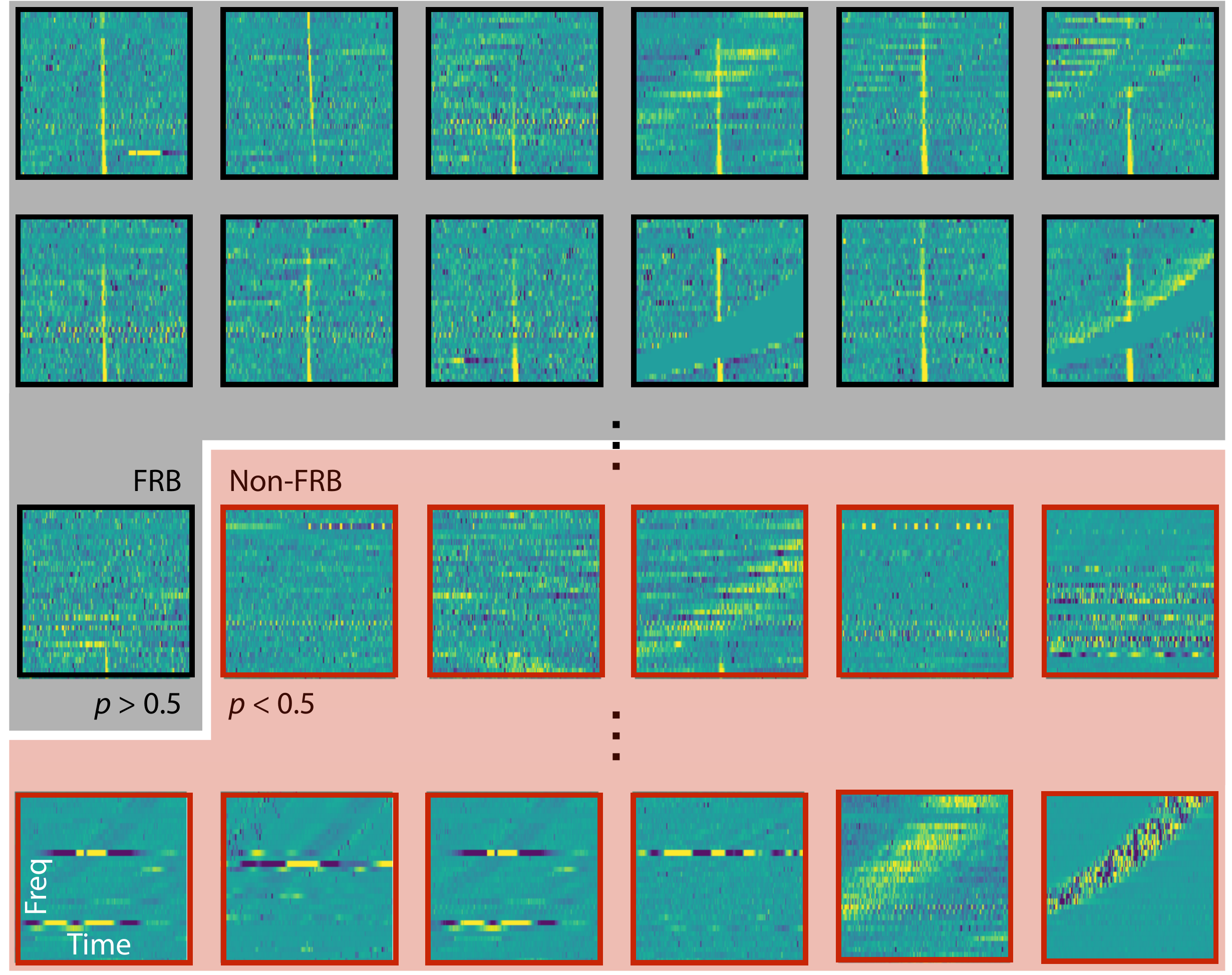}
	\caption{The ability of a model trained on 
			 simulated bursts to correctly identify 
			 Galactic pulsars. Our DNN classifier's output is 
			 a list of triggers ranked by their probability 
			 of being an FRB. Shown are frequency (ordinate) versus time (abscissa) 
			 arrays of the test triggers, identical to the top panels in Fig.~\ref{fig-triggers}.
                         The set includes real pulsars. 
			 Events that the 
			 classifier thinks are FRBs ($p>0.5$) are boxed in black, and 
			 non-FRBs are boxed in red. The most likely events 
			 are the top two rows, the marginal events are in 
			 the middle row, where 
			 the predicted labels 
			 transition from `FRB' to `RFI', and the final row 
			 are the least likely to be a true-positive. 
			 In this case we trained
			 on 4850 known false positives from the CHIME Pathfinder, 
			 and 4850 simulated FRBs. The test data is a separate 
			 set of several hundred triggers consisting of 
			 known false positives, single pulses from B0329+54,
			 and giant pulses from the Crab. The Pathfinder classifier gets fewer 
			 than 1$\%$ wrong. Our classifier trained on Apertif data 
			 achieves $99.7\,\%$ recall.}
\label{fig-ranked}
\end{figure}

One must be cautious when including simulated events 
in a machine learning training set. For this reason we carried out tests in which 
classifiers trained on simulated FRBs were used to 
predict the labels of real single pulses from 
Galactic pulsars. We did this for both Apertif  
and CHIME Pathfinder data independently, using their respective  
hand-labelled false positives in combination with simulated bursts
to train their DNNs. 
The models were then used 
on separate datasets containing hundreds of Crab giant pulses and 
B0329+54 single pulses. In Fig.~\ref{fig-ranked} we 
show the output of our pipeline for the CHIME Pathfinder dataset. 
For the CHIME Pathfinder, a recall of $\sim$\,99$\%$
can be achieved. Our Apertif model was trained 
on $\sim$\,20,000 candidates and applied to several 
hundred Galactic single pulses. It was able to 
recover 99.7\,$\%$ of these events.

\subsection{Speed}

For a given processor, training is always slower 
than classification since evaluation 
only requires one forward propagation 
through the neural network. For the applications 
described in previous sections, neither is 
prohibitively slow, even on CPUs. This is due to 
the modest sizes of the neural networks we have used (described in 
Sec.~\ref{sect-arch}). However, for real-time FRB surveys
like Apertif, ASKAP, UTMOST, and CHIME, it may be necessary to 
`decide' on triggers quickly and without human inspection, for example if voltages 
are to be saved, or if an alert is to be sent 
to another telescope.
Therefore, low-latency classification will be required.
Our measurements of executions times for training and classification,
shown in Fig.~7, 
indicate both can easily be done in 
real time on either CPUs and GPUs.
A single GTX Titan X can classify 10$^4$ candidates in under a second. 
For Apertif, the ARTS cluster contains 164 GTX 1080 Ti GPUs, which are each about twice as fast.
Each Apertif compound beam (Fig.~\ref{fig-beams}) is reduced on a dedicated 4-GPU server. The dedispersion and detection
routines \citep[{\tt AMBER}\footnote{\url{https://github.com/AA-ALERT/AMBER}};][]{2016A&C....14....1S} require the usage of only 2 of these. After RFI mitigation, 
the trigger levels in the single-pulse S/N can be set in {\tt AMBER}. Allowing of order 10 candidates to be are marked as
interesting, per second and per compound beam, would amount to $>$10 million candidates per day; among which may be of
order a single FRB. Even this liberal false-positive strategy could be easily further classified by the hybrid network,
running on a single GPU on the central ARTS server and VOEvent issuer.

\begin{figure}
    \includegraphics[clip,trim={0in, 0.1in, 0.8in, 0.9in}, width=\columnwidth]{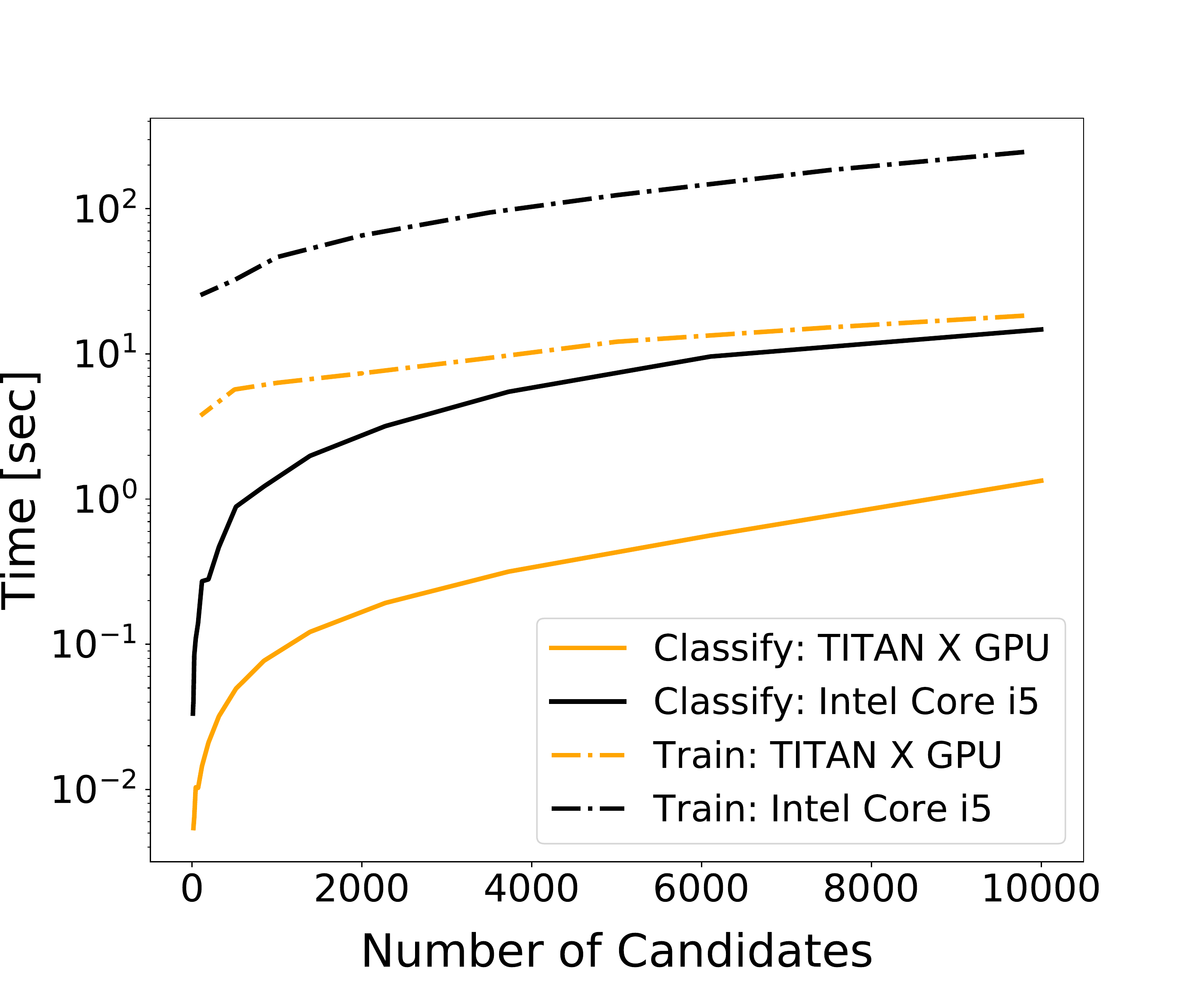}
	\caption{Classification and training time 
		     as a function of the number of dynamic spectra candidates. 
		     On a GTX Titan X GPU,
		     classification takes $\sim$\,100\,$\mu$s
		     per frequency-time array, using the network architecture 
		     described in Sec.~\ref{sect-arch} with the {\tt TensorFlow} 
		     backend of {\tt keras}. On a 2.9 GHz Intel Core i5-5287U CPU the classification 
		     time is a few milliseconds per candidate.
		     }	
\label{fig-speed}
\end{figure}

\subsection{Phased array feed simulation}
\begin{figure}
  \centering
  \includegraphics[width=0.75\columnwidth]{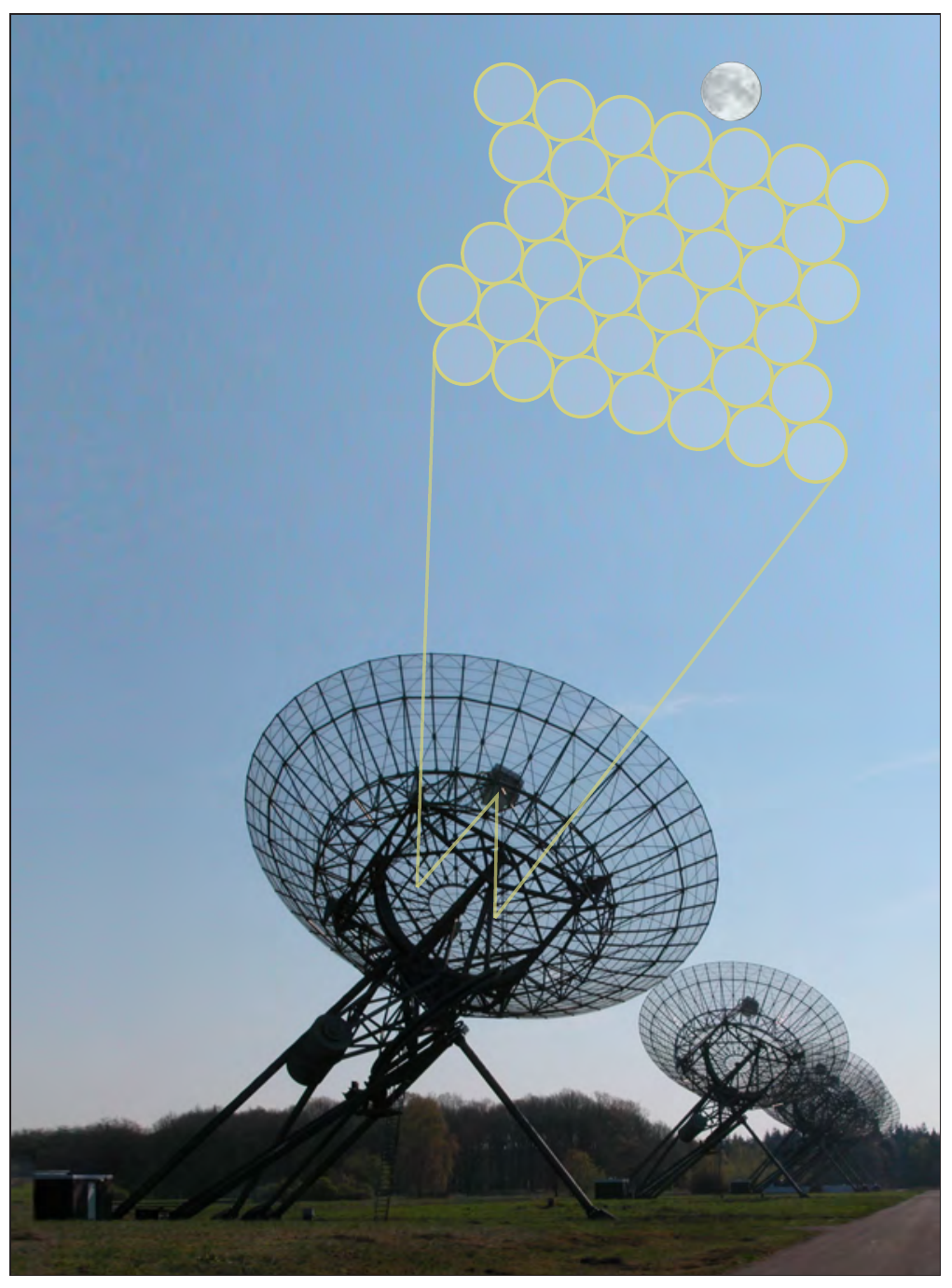}
	\caption{Lay out of the
          39 compound beams that can
          maximally be formed at full bandwidth. This pattern provides the most uniform
          sensitivity over the celestial sphere possible  (K.~Hess, \emph{priv.~comm.}).
		     }	
\label{fig-beams}
\end{figure}

To test the efficacy of including multi-beam detection information
in our DNN, we simulated  
the on-sky response within the 39 compound beams in the Apertif
phased array feed (Fig.~\ref{fig-beams}).
We randomly scattered 10,000 FRBs within this multi-beam setup, currently
 planned for the imaging and time-domain surveys with
 Apertif\footnote{\hspace{-1mm}\url{http://www.astron.nl/radio-observatory/apertif-surveys}}. 

Events were drawn from a Euclidean 
flux distribution, and ``detections'' greater than 6\,$\sigma$ were 
recorded. 
Though the fraction of multi-beam 
detections depends on the FRB brightness distribution, we found
no significant differences when reasonable non-Euclidean values were used.
RFI was assumed to show up in a random number of 
beams ranging from 1 to 39, following a log-normal distribution 
such that $75\%$ of events are in detected in 5 or more beams. 
The training set was then taken to 
be 15,000 simulated length-39 S/N vectors. The feed-forward 
neural network is then tested on the remaining 5,000 events, 
with an accuracy of $\sim$\,85$\%$ and a slightly higher 
recall, when using the described configuration. This is much 
worse than the dynamic spectra CNN (accuracy above $99\%$), 
but that is to be 
expected. Seeing an event in only one beam does not preclude 
its being false positive, in the same way multi-beam detection 
does not guarantee that the event was RFI. The multi-beam data 
add orthogonal, complementary information to pulse shape
and frequency structure, and was shown to be a valuable part of the hierarchical hybrid neural network (bottom row of Fig.~\ref{fig-tree}).

\section{Real-time classification with AI}
\label{sect-realtime}

Advances in signal processing allow the data volumes of future radio surveys to grow at considerable rates.
As the same evolution will also generally permit hardware to keep up, parallelized versions of existing software tools
may continue to analyze these data streams in real time \citep[cf.][]{2017arXiv171201008L}.
But this data deluge will likely out-pace the ability of the end-user astronomers to study all results.
Interferometers like 
the Square Kilometer Array \citep[SKA;][]{2009A&A...493.1161S}
and the next generation Very Large Array (ngVLA)\footnote{\url{http://ngvla.nrao.edu/}}
will not be able to search the newly available regions of parameter 
space simply by increasing person power---new techniques must 
be developed. It is therefore reasonable to ask how advances 
in machine learning might aid this endeavour. 

We investigate whether a real-time 
DNN classifier could be used for transient detection.
For example, might it be possible to completely replace dedispersion 
backends with a pre-trained neural network?
Even though the classifier could learn to identify arbitrary signals 
and not just $\nu^{-2}$ sweeps,
comparing to dedispersion provides a useful benchmark.

We can start by checking if any deep classifier could 
even keep up with the high data rates involved in 
real-time transient detection.
The data rates 
on modern multi-pixel radio telescopes are enormous, meaning 
data must either be searched in real-time or binned down to 
lower resolution for offline processing.
The scale of this challenge is exemplified by ARTS, the Apertif Radio Transient System
\citep{leeu14}. Within each of the 
 39 compound beams (cf.~Fig.~\ref{fig-tree}), ARTS forms 12 tied-array beams for full sensitivity \citep{2017arXiv170906104M}, whenever the telescope is active. 
This continuously produces 225\,Gbps of 300-MHz, 41\,{$\mu$}s data that needs to be searched in real-time.
A massive dedicated cluster with 164 GPUs (GTX 1080 Ti) is required to keep up with this data rate.

In the case of real-time classification of FRBs,
the idea would be to train on large numbers of dispersed pulses, 
such that the model would learn to look for $\nu^{-2}$ sweeps,
independent of their DM and with enough translational 
invariance to be insensitive to arrival times. This would 
supplant the need for dedispersion back-ends, which 
calculate a S/N after collapsing in frequency for multiple trial 
DMs. 

In the past, offline processing and 
lower data rates meant brute-force dedispersion was sufficient. 
The brute-force algorithm requires summing $N_f$
frequency channels for $N_{DM}$ DM trials for all $N_t$ time samples.
Its computational complexity is $\mathcal{O}(N_f N_t N_{DM})$
\citep{magro2011, barsdell2012, sclocco2014}. 
Tree dedispersion applies a divide-and-conquer technique by 
taking advantage of the redundancy in dedispersion for nearby 
frequency channels \citep{taylor74}. 
By using an FFT-like approach, the problem 
is reduced to a tree with $\log_2 N_f$ branches, allowing for
a $\mathcal{O}(N_f N_t \log_2 N_f)$ complexity. A highly optimized 
CPU-based version of this algorithm has been implemented 
for CHIME's FRB search \citep{chimbo}.
Other algorithms like the fast DM transform (FDMT)
exploit the same redundancy and attempt to maintain 
optimality \citep{zackay2014}. This algorithm is used 
in ASKAP's FREDDA pipeline \citep{bannister2018}. 

For a neural network like the one shown in 
Fig.~\ref{fig-CNN2d}, forward propagation is simply a 
series of convolutions and matrix multiplications. The two computational 
bottlenecks are the input layer, in which the $N_f$$\times$$N_t$ array 
is convolved with $n_{k_1}$ kernels, and the first fully-connected 
layer. A fully connected layer with $n$ inputs and $m$ outputs
scales as $\mathcal{O}(nm)$. This is because the output is given 
by,

\begin{equation}
\mathbf{z} = \mathbf{W}\mathbf{x} + \mathbf{b}
\end{equation}

\noindent where $\mathbf{x}$ is the $n$-element input vector,
$\mathbf{b}$ is a vector containing offsets of the $m$ neurons in 
that layer, and $\mathbf{W}$ is an $m$$\times$$n$ matrix 
whose elements $w_{ij}$ give the connection between the 
$j^{th}$ input and the $i^{th}$ neuron. 

In the final convolutional layer, $n_{k_2}$ arrays 
are created, one for each kernel in the second convolution.  
The final pooling step takes these $n_{k_2}$ matrices 
and reduces them in size by a factor of $p_x p_y$, by 
mapping each box of dimensions $p_x$ by $p_y$ 
to a single pixel is the subsequent layer. Therefore,
the input of the first fully-connected layer is an 
unravelled vector of length 

\begin{equation}
n_l = \frac{N_f N_t}{p_x^2 p_y^2} n_{k_2},
\end{equation}

\noindent since the original $N_f$$\times$$N_t$ array 
has been reduced in size twice by a factor of 
$p_x p_y$ through pooling. With $n_{d_1}$ neurons in 
the first fully-connected layer, calculating the 
activations of this component scales as, 

\begin{equation}
\mathcal{O} \left ( \frac{N_f N_t}{p_x^2 p_y^2} n_{k_2} n_{d_1} \right ).
\end{equation}

\noindent This means if the network's parameters are such that 

\begin{equation}
\frac{n_{k_2} n_{d_1}}{p_x^2 p_y^2} < N_{DM}
\label{eq-ineq}
\end{equation}

\noindent then this layer can be computed faster than brute-force 
dedispersion. In the model we used for classification of already-dedispersed 
single pulses, $p_x$ and $p_y$ were both 2, $n_{k_2}=64$ kernels 
were used in the last convolutional layer, and the first 
fully-connected layer had $n_{d_1}=128$ neurons. Thus, the inequality 
in Eq.~\ref{eq-ineq} would be satisfied for most brute-force searches, 
since the left side in our current model would be 512, 
wheras $N_{DM}\sim10^{3-4}$.

For dedispersion algorithms that are more optimized, such as subband or tree dedispersion, the balance in Eq.~\ref{eq-ineq} may be different; for tree dedispersion, the right-hand side term is $\log_2 N_f$, which is of order 10. So in contrast to the brute-force approach its computational intensity may be less than the DNN. 
That does not immediately imply, however, that the real-life performance of these optimized dispersion algorithms is proportionally faster. Dedispersion is a \emph{memory} bound algorithm for real-world parameters. Through data reuse, brute force dedispersion can approach the execution time of the more optimized algorithms \citep{2016A&C....14....1S}. Yet the fact that the matrix multiplications underlying the DNN are \emph{compute} bound can give the classifier a further real-life advantage on compute-biased accelerators such as GPUs.

Forward propagation through the 
first layer of our CNN amounts to computing $n_{k_1}$ convolutions. 
Convolution can be slow for two arrays of similar size. Using the 
brute-force method, this operation scales as $\mathcal{O}(N^{2D})$,
where $N$ is the input array's length and $D$ is number of dimensions. 
By invoking the convolution theorem, FFTs allow for a speed up, scaling as 
$\mathcal{O}(N^{D}\log_2^D N)$. However, our case is different from these, 
since our first layer requires convolving an $N_f$$\times$$N_t$ array with 
a much smaller array, often with kernels of size 3$\times$3
or 5$\times$5. The convolutions can be lowered to matrix multiplications, 
which are highly optimized on GPUs, allowing routines like 
{\tt cuDNN} and {\tt cuda-convnet2} high arithmetic intensity and efficiency 
\citep{cuDNN2014}. If we have a filter 
tensor that consists of $n_{k_1}$ kernels of size $n\times n$, 
then that can be reshaped to an array of dimensions $n_{k_1}\!\times n^2$. 
With batches of $N_b$ data arrays, 
and each of whose images are $N_f$$\times$$N_t$, then the data 
tensor can be reshaped to an $n^2\times N_b N_f N_t$ matrix. 
The convolution can then be computed as a matrix multiplication, 
which scales as,

\begin{equation}
\mathcal{O}(n_{k_1} n^2 N_b N_f N_t).
\end{equation}

\noindent Therefore each frequency-time array takes on average 
$n_{k_1} n^2 N_f N_t$ computations after dividing out the 
number of arrays per batch. In our case, with
16 or 32 length-3 square kernels, our most expensive 
convolutional layer is faster than brute-force dedispersion 
since $n_{k_1} n^2\approx 10^2 < N_{DM} \approx 10^4$. 
There are further techniques that 
allow a large DNN to be approximated by a smaller one. 
The deeper and/or wider model would 
be trained offline, and its compactified version could be 
applied in real-time classification 
at a faster speed but with similar accuracy. 

More than purely its speed, the sensitivity an algorithm provides is highly important when aiming to discover weak sources.
Thus, despite the somewhat surprising fact that a moderate
deep convolutional neural network could 
search raw intensity data faster than the brute-force 
dedispersion algorithm, we argue that dedispersion 
is not an ideal problem for deep learning. This is because 
algorithms like brute-force dedispersion, the FDMT, 
and tree-dedispersion are either optimal, or near-optimal
in signal recovery. With a 2D CNN of only a dozen layers, 
the model is more successful if S/N does not fall significantly 
below $\sim$\,1 per pixel. The universal approximation theorem states
that a finite feed-forward neural network can approximate arbitrary 
functions, meaning a sufficiently large network could, 
in principle, mimic optimal dedispersion \citep{cybenko1989}.
However, the theorem 
says nothing about such a network 
being reasonably sized, nor about its learnability. 
Still, by demonstrating the low theoretical complexity 
of classification, the radio community can consider the 
problems for which real-time deep learning classifiers 
might be suited. 
In the following section we discuss this further.

\section{Discussion}
\label{sect-discussion}

In this work we have found it sufficient to simulate FRBs
based on several parameters drawn from wide distributions. 
However if one wanted to improve the realism of 
true-positives in one's training set, 
there are new techniques that can be employed. 
Generative adversarial networks (GANs) are a class of deep learning 
algorithms that could generate realistic FRB candidates. They 
consist of two adversarial networks: one that generates 
realizations, 
and another that attempts to discriminate real from simulated data 
\citep{goodfellow2014}.
The generator's goal is to ``fool" the discriminator, eventually resulting 
in a high error rate in classification. This has allowed 
for the creation of photorealistic images based on drawings 
\citep{shrivastava2016}. \citet{guo2017} found that a standard 
deep CNN hit a performance ceiling for pulsar searching using real 
pulsars, so they used a deep convolutional GAN to build a collection 
of candidates. If such techniques were developed further, they may 
be useful for generating simulated RFI. In general, RFI is very difficult 
to model, but with an adversarial network trained on unlabelled 
real data, the problem of parametrizing it by hand could be overcome.

The black-box problem is another general concern about 
using deep neural networks in place of more explicit modelling. 
While we consider this a genuine issue for other problems, 
in the case of false-positive sifting our multi-input artificial net
is no more opaque than a human scientist's biological neural network.
A human scientist knows some basic facts about dedispersed FRBs---they are 
roughly broad-band, narrow in time, etc.---and then gets a ``feel'' 
for what false-positives look like by inspecting $10^3$--$10^4$ triggers. 
We never really know which features the expert has deemed 
salient, whereas in Fig.~\ref{fig-CNN2d} we show the actual 
activations inside of our neural network for a given input. 
Therefore if our goal is simply to save time by
accurately filtering out false-positives, 
the black-box problem is not a major consideration.

Having a machine learning classifier that can keep up 
with real-time triggers will be useful for a number of reasons.
Even if all candidates are to be written to disk, 
the number of false positives may end up being prohibitively 
large for email notifications, outriggers, voltage dumps, 
or VOEvents. 
Because our neural network assigns a probability to each 
candidate, groups can set up a confidence threshold, below which 
triggers are saved but do not effect an alert. 
 
We have also discussed the possibility of not only sifting 
through high-significance dedispersed candidates in real-time, 
but actually searching raw data in place of dedispersion backends.
Beyond the optimized routines we described in 
Sec.~\ref{sect-realtime}, the training of, and classification with, 
deep neural nets is being made faster by tailored GPU 
hardware. Nvidia has released Tensor Cores in their 
Volta-based Tesla V100, which provide almost an 
order of magnitude speed up in matrix multiplication 
for large arrays over the Pascal-based P100 GPU.
Google has also responded to the increased use of DNNs 
by building a custom application specific integrated circuit (ASIC)
that they have called ``Tensor Processing Units'' 
(TPUs) \citep{tpu2017}.
These TPUs cannot yet help train neural networks, but 
were built specifically for classification, ideal for what we have described in 
Sec.~\ref{sect-realtime}.

We showed that the computational 
complexity of a single forward propagation through a
modest CNN can be significantly less than that of 
brute-force dedispersion. Furthermore, dedispersion 
algorithms tend to have low arithmetic intensity 
which means they are memory bound and not ideal 
for GPUs. Classification using neural networks 
amounts to a series of matrix multiplications, 
accelerated by previously discussed hardware. 
However, we argue that a 
CNN could not reach the level of statistical optimality 
of known dedispersion algorithms without 
making the network so large that gains in speed 
were lost. 

Applying deep learning to real-time 
transient detection may still be useful in upcoming surveys. 
Dedispersion algorithms search for signals 
with $\nu^{-2}$ sweeps, caused by the differential 
group velocity of light in cold dense plasmas. 
Deviations from such a quadratic dispersion relation 
can come from relativistic plasmas, or electrons 
whose plasma frequency is close to the observing frequency.
Unusual polarization signatures can be induced by propagation, 
which can be searched for
\citep{kennett1998}. SETI might also find these 
techniques useful in searching for  
bright, structured signals from 
extraterrestrial civilizations.
But history teaches us that 
the most exciting discoveries in transient astronomy 
come from ``unknown unknowns'', usually by searching a 
parameter space that was not previously accessible. 
The SKA and ngVLA will offer such datasets, and their 
availability may coincide with great advances in 
unsupervised, or semi-supervised learning. 

\section{Conclusions}

We have applied deep learning to the problem of 
single-pulse classification, with large real-time 
FRB surveys in mind. Using Google's {\tt TensorFlow} 
we developed a multi-input deep neural network that takes 
FRB candidate diagnostic data, such as dynamic spectra, 
the DM-time intensity array, and multi-beam information, and 
returns a probability of the event being real. Models 
can be trained offline but applied in real-time, allowing for 
low-latency classification if outriggers, VOEvents \citep{petroff2017VO}, 
or voltage dumps 
are to be triggered. 
These tools are available on 
github\footnote{\url{https://github.com/liamconnor/single\_pulse\_ml}}.

The possibility of replacing dedispersion backends with 
a single DNN classifier was investigated. 
Although statistical optimality to purely quadratically dispersed signals may not be achievable without 
cumbersome multi-layer models, we showed that  
forward propagation could be done more efficiently and 
quickly than brute-force dedispersion on modern hardware.
Thus, deep learning classification of signals more diverse than dispersion is feasible, on raw data, in real-time. 

\begin{acknowledgements}
We thank Emily Petroff for helpful 
comments on the manuscript.
We thank the CHIME Collaboration for allowing 
the usage of data from its Pathfinder instrument,
and the Apertif Survey Team for the use of the ARTS false-positive 
dataset.  
We also thank Jorn Peters, Yunfan (Gerry) Zhang, 
and Folkert Huizinga for useful discussions.
The research leading to these results has received funding from the European Research Council  under the European Union's Seventh Framework Programme (FP/2007-2013) / ERC Grant Agreement n. 617199, and from the Netherlands Research School for Astronomy (NOVA4-ARTS).
\end{acknowledgements}

\bibliography{dl_paper}
\bibliographystyle{yahapj}

\end{document}